\documentstyle[amssymb,preprint,aps]{revtex}

\newtheorem{theorem}{Theorem}

\newtheorem{definition}[theorem]{Definition}

\newtheorem{proposition}[theorem]{Proposition}
\newtheorem{remark}[theorem]{Remark}

\begin{document}
\title{Bloch waves and non-propagating modes in photonic crystals}
\author{R. Sma\^{a}li, D. Felbacq, G. Granet}
\address{LASMEA UMR-CNRS 6602\\
Complexe des C\'{e}zeaux\\
63177 Aubi\`{e}re Cedex\\
France}
\date{\today }
\maketitle

\begin{abstract}
We investigate the propagation of electromagnetic waves in finite photonic
band gap structures. We analyze the phenomenon of conduction and forbidden
bands and we show that two regimes are to be distinguished with respect to
the existence of a strong field near the interfaces. We precise the domain
for which an effective medium theory is sounded.
\end{abstract}

\bigskip \tightenlines

\section{Introduction}

The theoretical and numerical study of photonic band gap materials (see \cite
{web} for an exhaustive bibliography) may be led both from the point of view
of the spectrum, with the computation of the band structure by using Bloch
wave theory and the related quantities (Density of States and so on), or
from the point of view of the scattering theory \cite{moi}. In case of a
scattering experiment, which of course involves a finite structure, Bloch
waves are not {\it a priori} sufficient to describe the electromagnetic
field because of the existence of evanescent and anti-evanescent waves,
linked to the boundary of the device. Of course these non-propagating modes
are of prime importance in the band gaps, as they represent entirely the
total field inside the photonic crystal, but they may also induce strong
effects for frequencies lying inside the conduction bands.

In the context of electromagnetic optics and the spectacular effects that
can be obtained using photonic crystals (the so-called ultra-refraction
effects \cite{dowling,kosaka,gralak,maystre1,ultra,motoni}) a theoretical
approach only involving equivalent medium theories, group velocity \cite
{dowling,kosaka,gralak} and more generally quantities only derived from the
band structure is certainly incomplete. Previous studies have shown the
importance of considering the isoenergy diagram and not only the dispersion
diagram \cite{motoni,branch}. In the present work, our aim is to analyse the
relative importance of the evanescent waves (the near-field) for describing
the electromagnetic field inside a photonic crystal. More precisely, we
study the ratios of the projection of the field on the propagating and
non-propagating modes, and we precise the conditions under which Bloch waves
can describe entirely the scattering behaviour of a photonic crystal.

\section{Propagating and non-propagating modes}

The photonic crystal under study is made of a stack of gratings. For the
numerical applications we use square dielectric rods in a dielectric matrix
(fig. 1). The basic layer of the device is made of one grating, consisting
of a slab of rods between two homogeneous slabs above and below (fig.1): it
covers a band ${\Bbb R\times }\left[ -h/2,h/2\right] $. The period of the
grating is $d$. We use time-harmonic fields with a time dependence of $%
e^{-i\omega t}$ and $k=\frac{2\pi }{\lambda }$ denotes the wavenumber in
vacuum. We have chosen this particular geometry (square rods and rectangular
symmetry) and polarization for the sake of simplicity (both on the numerical
level and for the theoretical exposition). However the theoretical framework
is quite general and works as well for more complicated situations (see \cite
{pend1} for an exposition of the transfer matrix method for 3D obstacles).

Our aim is to characterize the field inside the photonic crystal. Due to the
translational invariance of the problem, we look for pseudo-periodic fields
in the $x$ direction: $u\left( x+d,y\right) =\exp \left( i\alpha d\right)
u\left( x,y\right) $, where $\alpha $ belongs to the interval $Y=\left[ -\pi
/d,\pi /d\right[ $. In a diffraction problem, where the grating is
illuminated by a plane wave under the incidence $\theta $ (fig. 1), we have $%
\alpha \equiv k\sin \theta $ $%
\mathop{\rm mod}%
\left( \pi /d\right) $. From grating theory \cite{yellow}, we can define the
transfer matrix of the basic layer, which is the operator $\widetilde{{\bf T}%
}_{\left( \alpha ,\lambda \right) }$ linking $\left( u(x,h/2),\partial
_{y}u\left( x,h/2\right) \right) $ to $\left( u\left( x,-h/2\right)
,\partial _{y}u\left( x,-h/2\right) \right) $ (in ordinary differential
equations theory this is the monodromy matrix). There are many very good
numerical methods for computing the field diffracted by a grating and hence
the transfer matrix, even for stacks of gratings \cite{pend2,granet,li}.

On the upper (resp. lower) side of a basic cell, we expand the field on a
Rayleigh basis: 
\begin{equation}
u(x,h/2)=\sum_{n}\left( A_{n}^{+}+A_{n}^{-}\right) e^{i\alpha
_{n}x},u(x,-h/2)=\sum_{n}\left( B_{n}^{+}+B_{n}^{-}\right) e^{i\alpha _{n}x}.
\end{equation}
\newline
where $\alpha _{n}=\alpha +nK,K=\frac{2\pi }{d}$. The values of the normal
derivatives $\left. \partial _{y}u\right| _{y=\pm h/2}$ write: 
\[
\partial _{y}u(x,h/2)=\sum_{n}i\beta _{n}\left( A_{n}^{+}-A_{n}^{-}\right)
e^{i\alpha _{n}x},\partial _{y}u(x,-h/2)=\sum_{n}i\beta _{n}\left(
B_{n}^{+}-B_{n}^{-}\right) e^{i\alpha _{n}x}. 
\]
The knowledge of $\widehat{A^{\pm }}=\left\{ A_{n}^{\pm }\right\} _{n}$
(resp. $\widehat{B^{\pm }}=\left\{ B_{n}^{\pm }\right\} $) gives the value
of the derivatives. Therefore, rather than computing the monodromy matrix as
defined above, we compute the matrix ${\bf T}_{\left( \alpha ,\lambda
\right) }$ such that 
\begin{equation}
{\bf T}_{\left( \alpha ,\lambda \right) }\left( 
\begin{array}{c}
\widehat{A^{+}} \\ 
\widehat{A^{-}}
\end{array}
\right) =\left( 
\begin{array}{c}
\widehat{B^{+}} \\ 
\widehat{B^{-}}
\end{array}
\right)
\end{equation}

The point is to analyze the spectrum of ${\bf T}$. For symmetry reasons, the
spectrum $sp\left( {\bf T}\right) $ of ${\bf T}$ is invariant under $\tau
\rightarrow \tau ^{-1}$(this is easily seen in case of a lamellar grating,
but the proof is slightly more involved in case of a $y$-dependent medium),
then we can distinguish between eigenvalues of modulus $1$ that are
necessarily finitely many, and eigenvalues that do not belong to the unit
circle of the complex plane. Let us denote $e^{i\beta h},\beta \in \left] -%
\frac{\pi }{h},\frac{\pi }{h}\right] ,$ an eigenvalue of ${\bf T}_{\left(
\alpha ,\lambda \right) }$ of modulus one, and $\psi $ an associated
eigenvector then we have ${\bf T}\left( \widehat{\psi ^{+}},\widehat{\psi
^{-}}\right) =e^{i\beta h}\left( \widehat{\psi ^{+}},\widehat{\psi ^{-}}%
\right) $ or else: $\psi \left( x,y+h\right) =e^{i\beta h}\psi \left(
x,y\right) $ this means that $\left( x,y\right) \rightarrow e^{i\alpha
x}\psi \left( x,y\right) $ is a Bloch wave associated to the Bloch vector $%
\left( \alpha ,\beta \right) $. That way we can easily compute the
dispersion curves at a given wavenumber $k$. Moreover, we can also compute
the non-propagating modes inside the crystal: they correspond to
eigenvectors associated with eigenvalues that are not of modulus one. We
have thus obtained a decomposition of the modes by means of a family of
monodromy operators parametrized by $\alpha \in Y$. As it as been said
before, from the scattering point of view, the parameter $\alpha $ is equal
to $k\sin \theta $ so that we study the spectrum of ${\bf T}_{\left( \theta
,\lambda \right) }={\bf T}_{\left( k\sin \theta ,\lambda \right) }$.

We can now give the following definitions.

\begin{definition}
We call {\bf relative gap} an interval of wavelengths $I_{\theta }$, at a
given incidence $\theta $, for which ${\bf T}_{\left( \theta ,\lambda
\right) }$ has no eigenvalues of modulus one and we call {\bf relative
conduction band} an interval $B_{\theta }$ of wavelengths where ${\bf T}%
_{\left( \theta ,\lambda \right) }$ does have eigenvalues of modulus $1$.
\end{definition}

\begin{definition}
A {\bf total gap} corresponds to the intersection of incident dependent gaps
(and may be void).
\end{definition}

\section{Analysis of the spectrum}

As it has already been stated, at a given wavelength, Bloch waves are not
sufficient to compute the scattering properties of the crystal. In order to
quantify the relative importance of the evanescent waves, we need to be able
to get a decomposition of the field.

\subsection{Decomposition of the field}

Once the electromagnetic field is known on the upper face of the crystal
(through the coefficients $\widehat{A^{\pm }}$ , obtained from a rigorous
numerical method), it is possible to expand it on the various modes that
exist in the grating layer. More precisely, except on a set of wavelengths
of zero Lebesgue measure, matrix ${\bf T}_{\left( \alpha ,\lambda \right) }$
can be put in diagonal form: 
\[
{\bf T}_{{\bf \left( \alpha ,\lambda \right) }}{\bf =T}_{p}{\bf \oplus T}_{e}%
{\bf \oplus T}_{a} 
\]
where ${\bf T}_{p}$ is a finite rank operator corresponding to propagative
waves and ${\bf T}_{e}${\bf ,}${\bf T}_{a}$ correspond to the evanescent and
anti-evanescent modes.

With this decomposition, the vector $\psi =\left( \widehat{\psi }^{+},%
\widehat{\psi }^{-}\right) $ writes $\psi =\psi _{p}\oplus \psi _{e}\oplus
\psi _{a}$. Whence we define the branching ratios\ $\pi _{p}$ (resp. $\pi
_{e},$ $\pi _{a}$) of the field on the propagating (resp. evanescent,
anti-evanescent) modes by: 
\begin{equation}
\pi _{p}=\frac{\left\| \psi _{p}\right\| ^{2}}{N\left( \psi \right) },\pi
_{e}=\frac{\left\| \psi _{e}\right\| ^{2}}{N\left( \psi \right) },\pi _{a}=%
\frac{\left\| \psi _{a}\right\| ^{2}}{N\left( \psi \right) }
\end{equation}
where $N\left( \psi \right) =\left\| \psi _{p}\right\| ^{2}+\left\| \psi
_{e}\right\| ^{2}+\left\| \psi _{a}\right\| ^{2}$.

The point of the above decomposition is to quantify the relative importance
of the various modes in the total field existing in the crystal, in order to
understand to what extend the field is not solely described by Bloch waves.

\subsection{Cut wavelengths and classification of the conduction bands}

Let us now turn to some numerical computations. The relative permittivity of
the rods is $\varepsilon _{2}=9$ and $\varepsilon _{1}=\varepsilon _{ext}=1$%
, the geometric parameters are $h=2.8,d=2.8,h_{1}=1.9$, $d_{1}=1$. The
structure is made of one basic layer and we choose $\theta =30%
{{}^\circ}%
$ and $s$-polarized waves. In fig.2 (a), we give the absolute values of the
eigenvalues of ${\bf T}_{\theta }\left( \lambda \right) $ versus the
wavelength. The conduction bands are the regions with a horizontal straight
line ($\left| 
{\mu}%
\right| =1$). For each wavelength $\lambda $ there is a finite, possibly
empty, set of eigenvalues of modulus one $\left\{ e^{i\beta _{n}\left(
\lambda \right) }\right\} _{n}$ and an infinite set of eigenvalues that do
not belong to ${\Bbb U=}\left\{ z\in {\Bbb C},\left| z\right| =1\right\} $
(in fig. 2 (b) we have plotted the real part of the spectrum of ${\bf T}%
_{\left( \theta ,\lambda \right) }$, where the finite number of propagating
modes may be observed). We have also plotted in fig.2 (c) the dispersion
diagram $\left( \beta ,\lambda /d\right) $. The comparison with fig. 2 (a)
shows that the consideration of the complete spectrum of ${\bf T}_{\left(
\theta ,\lambda \right) }$ , i.e. with the non-propagating modes, allows to
understand that one should not treat on a different foot propagating and
non-propagating modes because they are really the same physical entities,
behaving differently according to the wavelength. When $\theta $ varies in $%
\left] -\pi /2,\pi /2\right[ $, the local gaps vary as shown in fig. 3.

This first example is rather generic and shows that within a given
conduction band, hence {\it locally}, it is possible to define continuous
sections $\lambda \rightarrow \mu _{n}\left( \lambda \right) \in sp\left( 
{\bf T}_{\left( \theta ,\lambda \right) }\right) $ representing the
evolution of the eigenvalues of the monodromy operator with respect to the
wavelength. At some values of the wavelength however, these sections may
encounter a bifurcation, or cut-off: the eigenvalue leaves ${\Bbb U}$ and
the associated modes give rise to an evanescent mode and an anti-evanescent
mode. At such a branch point, the section $\lambda \rightarrow \mu
_{n}\left( \lambda \right) $ is not differentiable and may cross other
sections. As a consequence, a global description of the sections is not
possible in that case. This problem is quite a complicated one, for which
there is a general theory \cite{kato,geophys}, but even deriving a specific
theory for our particular situation is quite a big task and beyond the scope
of this work.

However, an easy simplification can be obtained by noting that the set of
eigenvalues being invariant under $\tau \rightarrow \tau ^{-1}$, it is
natural to consider the quotient space $sp\left( {\bf T}_{\left( \theta
,\lambda \right) }\right) /\sim $ for the equivalence relation $\mu _{1}\sim
\mu _{2}$ if $\mu _{1}\mu _{2}=1$, which amounts to identify two eigenvalues
that are inverse one of the other. This operation gives a nice
simplification but still does not allow to define global sections (a more
detailed account of this situation will be given elsewhere \cite{moi2}).

\begin{remark}
A very simple example of branch point is the extinction of a diffracted
order in grating theory. Another elementary situation is that of a
stratified medium (a Bragg mirror for instance) in which case there are only
two propagative modes in the conduction bands and one evanescent and one
anti-evanescent mode inside the gaps. A realization of the quotient space is
obtained by considering $\left\{ \frac{1}{2}tr\left( {\bf T}_{\left( \theta
,\lambda \right) }\right) ,\lambda \in {\Bbb R}^{+}\right\} $. In that case,
this very set defines a global section and the quotient space $sp\left( {\bf %
T}_{\left( \theta ,\lambda \right) }\right) /\sim $ is a trivial fibred
bundle.\newline
\end{remark}

For a given incidence, a gap is then an interval of wavelengths over which
all the propagative eigenvalues have encountered a bifurcation. In fig. 2
(a), we have this situation in the interval $\left( 1.32,1.42\right) $.

When the wavelength tends to infinity, it is known that the device finally
behaves as a homogeneous slab \cite{lekner,wavepropag,kozlov,homog}, and
then there are only two propagative modes (up and down), which means that
all ''sections'' finally bifurcate definitely (see fig. 2 (a-b) for $\lambda
/d>2.22$), except the one corresponding to the homogenization regime. In
that case there are still evanescent (and anti-evanescent waves) but with a
very huge damping exponent so that $\pi _{e}$ and $\pi _{a}$ are small.

However, before that regime, eigenvalues may experience a local bifurcation:
that is they leave ${\Bbb U}$ over a finite interval but finally come back
on it (in fig. 2 (b) this situation happens\ over the interval $\left(
1.63,1.87\right) $). What is important to note is that such a local
bifurcation may affect only one eigenvalue so that, whithin a conduction
band there may be evanescent field coming from such a bifurcation, hence
with a small damping exponent (this happens over the interval $\left(
1.32,1.42\right) $ in fig. 2 (a-b)). This leads us to distinguish between
both regimes and give the following definitions.

\begin{definition}
A conduction band is said {\bf local} if among the evanescent modes within
this band there is at least one mode corresponding to a local bifurcation. A
conduction band is {\bf global} if it is not local.
\end{definition}

Let us now give some numerical examples of the various regimes described
above. We give in fig.4 the absolute values of the eigenvalues of matrix $%
{\bf T}_{\left( 0,\lambda \right) }$ (normal incidence) for a one layer
structure with $\varepsilon _{ext}=2.26,\varepsilon _{1}=1,\varepsilon
_{2}=4 $ $h=1,d=1,d_{1}=0.5$ and the projection ratii $\pi _{p,e,a}$. The
region $\left( 1.28,1.37\right) $ corresponds to a local conduction band,
i.e. in which there is a local bifurcation of an eigenvalue. We see that the
part of the field on the non-propagating modes is not at all negligible so
that the field cannot be described solely by Bloch modes. On the contrary,
for the interval $\left( 1.72,1.8\right) $ the conduction band corresponds
to a global bifurcation of two eigenvalues of modulus one and in that case
the damping exponents are exponentially growing, so that almost Bloch waves
only contribute to the description of the field.

A natural question is to know to what extend these situations persist when
the number of layers is increased. We have computed the values of the
branching ratios when the number $N$ of layers is $N=2,4,6$. The results are
given in figures 5 (a-b-c). It can be seen that the branching ratios $\pi
_{e}$ and $\pi _{p}$ exhibit an oscillatory behavior with respect to $%
\lambda $, which is probably linked to the excitation of resonances, but
that the fraction of electromagnetic energy that is carried by the
evanescent waves is not diminished. This means that a non negligible part of
the field is localized near the interfaces, which can have substantial
consequences on the propagation of a beam inside the structure (this
situation will be analyzed in a forthcoming paper\cite{wavepropag}) but also
on the local density of states for photons.

As an application, suppose now that the wavelength belongs to a global
conduction band and is such that there is only one diffracted and one
transmitted order; then from the knowledge of the transmission and
reflection coefficients, it is possible to compute the ($2\times 2$)
monodromy matrix of the device. In that case, we can obtain the superior
envelope of the transmitted energy by considering only the transfer matrix
of one basic layer \cite{strat}. We have shown in another article \cite
{jphys} that in a layer characterized by a $2\times 2$ transfer matrix, the
reflected and transmitted coefficients for $N$ layers can be obtained in
close form : 
\begin{equation}
r_{N}\left( \lambda ,\theta \right) =f%
{\displaystyle{\mu ^{2N}-1 \over \mu ^{2N}-fg^{-1}}}%
,\text{ }t_{N}\left( \lambda ,\theta \right) =\mu ^{N}%
{\displaystyle{\left( 1-fg^{-1}\right)  \over \mu ^{2N}-fg^{-1}}}%
\label{rt}
\end{equation}
\newline
where $\mu $ is an eigenvalue of ${\bf T}_{p}=\left( {\bf t}_{ij}\right) $
associated with an eigenvector ${\bf u}=\left( u_{1},u_{2}\right) $, an
eigenvector associated to $\mu ^{-1}$ is denoted by ${\bf v}=\left(
v_{1},v_{2}\right) $ and, denoting $q(x_{1},x_{2})={{\frac{i\beta
_{0}x_{2}-x_{1}}{i\beta _{0}x_{2}+x_{1}}}}$, $\beta _{0}=%
{\displaystyle{2\pi  \over \lambda }}%
\cos (\theta )$, functions $f$ and $g$ are defined by 
\begin{equation}
\begin{array}{ll}
\text{if }\left( k,\theta \right) \in {\bf G} & g\left( k,\theta \right)
=q\left( {\bf v}\right) ,\text{ }f\left( k,\theta \right) =q\left( {\bf w}%
\right) \\ 
\text{if }\left( k,\theta \right) \in {\bf B} & \left\{ 
\begin{array}{l}
g\left( k,\theta \right) =q\left( {\bf v}\right) ,f\left( k,\theta \right)
=q\left( {\bf w}\right) \hbox{ if }\left| q\left( {\bf v}\right) \right|
<\left| q\left( {\bf w}\right) \right| \\ 
g\left( k,\theta \right) =q\left( {\bf w}\right) ,f\left( k,\theta \right)
=q\left( {\bf v}\right) \hbox{ if }\left| q\left( {\bf w}\right) \right|
<\left| q\left( {\bf v}\right) \right|
\end{array}
\right.
\end{array}
\end{equation}
where ${\bf G=}\left\{ \left( k,\theta \right) ,\left| tr({\bf T}%
_{p})\right| >2\right\} $ and ${\bf B=}\left\{ \left( k,\theta \right)
,\left| tr({\bf T}_{p})\right| <2\right\} .$\newline

\noindent The superior envelope $R_{\infty }$ of the reflected energy, and
conversely the inferior envelope of the transmitted energy $T_{\infty }$ are
given by \cite{strat}: 
\begin{equation}
\begin{array}{l}
T_{\infty }=%
{\displaystyle{4-tr\left( {\bf T}_{p}\right) ^{2} \over \left( {\bf t}_{12}\beta _{0}-{\bf t}_{21}\beta _{0}^{-1}\right) ^{2}}}%
\\ 
R_{\infty }=1-T_{\infty }
\end{array}
\label{epi}
\end{equation}
\newline
A direct application of these formulas show a very accurate result in fig.6
(a-b) for a global conduction band.

We can conclude by the following:

\begin{proposition}
Within a global conduction band, the field inside the crystal can be
represented by Bloch waves only.\newline
\end{proposition}

\section{Conclusion}

We have shown that it is important to distinguish between various kinds of
conduction bands: there may be non propagative modes that result from the
local bifurcation of a propagative mode or all the non propagative modes may
be made out of global bifurcation of propagative modes. In the first case,
an important part of the field inside the structure is made of
non-propagative modes, but in the second case, the field writes in terms of
Bloch waves only. Especially near a band edge, one should be very careful
before deriving the behavior of the field solely by looking at the
dispersion diagram: it does not take into account the evanescent waves.
These results might be useful in studying beam propagation and superprism
effects in photonic crystals\cite{ultraref} and also the phenomenon of
spontaneous emission in a finite dielectric structure, for in that case the
density of modes for photons derived without precaution from Bloch theory is
certainly false because the atom modes can couple to non-propagating
radiation modes \cite{moroz}.

\newpage

{\LARGE Figures captions}

Figure 1: Sketch of the photonic crystal.

Figure 2: (a) Absolute value of the eigenvalues of ${\bf T}_{\theta }\left(
\lambda \right) $ for $\varepsilon _{1}=9,\varepsilon _{2}=\varepsilon
_{ext}=1,h/d=1,h_{1}/d=0.68,d_{1}/d=0.35,s$-polarized waves.

\qquad \qquad (b) Real part of the spectrum of ${\bf T}_{\theta }\left(
\lambda \right) $. The arrows indicate a local and a global bifurcation.

\qquad \qquad (b) Dispersion diagram $\left( \beta ,\lambda /d\right) $ for
the parameters of figure 2 (a).

Figure 3: Evolution of the gaps with $\theta $. The white regions correspond
to band gaps.

Figure 4: Branching ratios for $\varepsilon _{ext}=2.26,\varepsilon
_{2}=4,\varepsilon _{1}=1,h/d=1,d_{1}/d=0.5,s$-polarized waves.

$\qquad \qquad \Box $: propagating ratio $\left( \pi _{p}\right) ,$ $%
\triangle $: evanescent ratio $\left( \pi _{e}\right) ,$ $\ast $:
anti-evanescent ratio $\left( \pi _{a}\right) .$

Figure 5: Spectrum and branching ratios for the parameters of figure 4.

\qquad \qquad (a) $N=2$ (b) $N=4$ (c) $N=6.$

Figure 6: Reflected energy and its enveloppe given by (\ref{epi}), with $%
N=20.$

\qquad \qquad (a) $\varepsilon _{1}=9,\varepsilon _{2}=\varepsilon
_{ext}=1,h/d=1,h_{1}/d=0.68,d_{1}/d=0.35,s$-polarized waves.

\qquad \qquad (b) $\varepsilon _{ext}=2.26,\varepsilon _{2}=4,\varepsilon
_{1}=1,h/d=1,d_{1}/d=0.5,s$-polarized waves.


\begin{references}
\bibitem{web}  http://home.earthlink.net/\symbol{126}jpdowling/pbgbib.html

\bibitem{moi}  E. Centeno, D. Felbacq, J. Opt. soc. Am. A 17 (2000) 320.

\bibitem{dowling}  J. P. Dowling, C. M. Bowden, J. Mod. Opt. 41 (1994) 435.

\bibitem{kosaka}  H. Kosaka et al., Phys. Rev. B 58 (1998) 10096.

\bibitem{gralak}  B. Gralak, G. Tayeb and S. Enoch, J. Opt. Soc. Am. A 17
(2000) 6.

\bibitem{maystre1}  S. Enoch, G. Tayeb and D. Maystre, Opt. Comm. 161{\bf \ }%
(1999) 171 .

\bibitem{ultra}  D. Felbacq, B. Guizal and F. Zolla, J. Opt. A: Pure Appl.
Opt. 2 (2000) L30 .

\bibitem{motoni}  M. Notomi, Phys. Rev. B 62 (2000) 10696.

\bibitem{branch}  T. Minami, H. Ajiki, K. Cho, Physica E 13 (2002) 432.

\bibitem{pend1}  J. B. Pendry, J. Mod. Opt. 41 (1994) 209.

\bibitem{yellow}  R. Petit ed., {\it Electromagnetic theory of gratings},
Springer-Verlag, Berlin, 1980.

\bibitem{pend2}  A. J. Ward, J. B. Pendry, W. J. Stewart, J. Phys.: Condens.
Matter 7 (1995) 2217.

\bibitem{granet}  G. Granet, J. Chandezon, O. Coudert, J. Opt. Soc. Am. A 14
(1997) 1576.

\bibitem{li}  L. Li, J. Opt. Soc. Am. A 13 (1996) 1024.

\bibitem{moi2}  D. Felbacq, in preparation.

\bibitem{lekner}  J. Lekner,{\it \ }J. Opt. Soc. Amer. A 11 (1994) 2892

\bibitem{wavepropag}  D. Felbacq, B. Guizal, F. Zolla, Opt. Comm. 152 (1998)
119.

\bibitem{kozlov}  V. Jikov, S. Kozlov, O. Oleinik, {\it Homogenization of
differential operators and integral functionals}, Springer-Verlag, Berlin,
1994.

\bibitem{homog}  D. Felbacq, G. Bouchitt\'{e}, Waves in Random Media 7
(1997) 245.

\bibitem{kato}  T. Kato, {\it Perturbation theory for linear operators},
Springer-Verlag, Tokyo, 1984.

\bibitem{geophys}  Th. Frankel, {\it The geometry of physics, }Cambridge
University Press, 1997.

\bibitem{strat}  D. Felbacq, B. Guizal, F. Zolla, J. Math. Phys. 39 (1998)
4604.

\bibitem{jphys}  D. Felbacq, J. Phys. A: Math. Gen. 33 (2000) 7137.

\bibitem{ultraref}  R. Sma\^{a}li, D. Felbacq, in preparation.

\bibitem{moroz}  A. Moroz, Europhys. Lett. 46 (1999) 419.
\end{references}
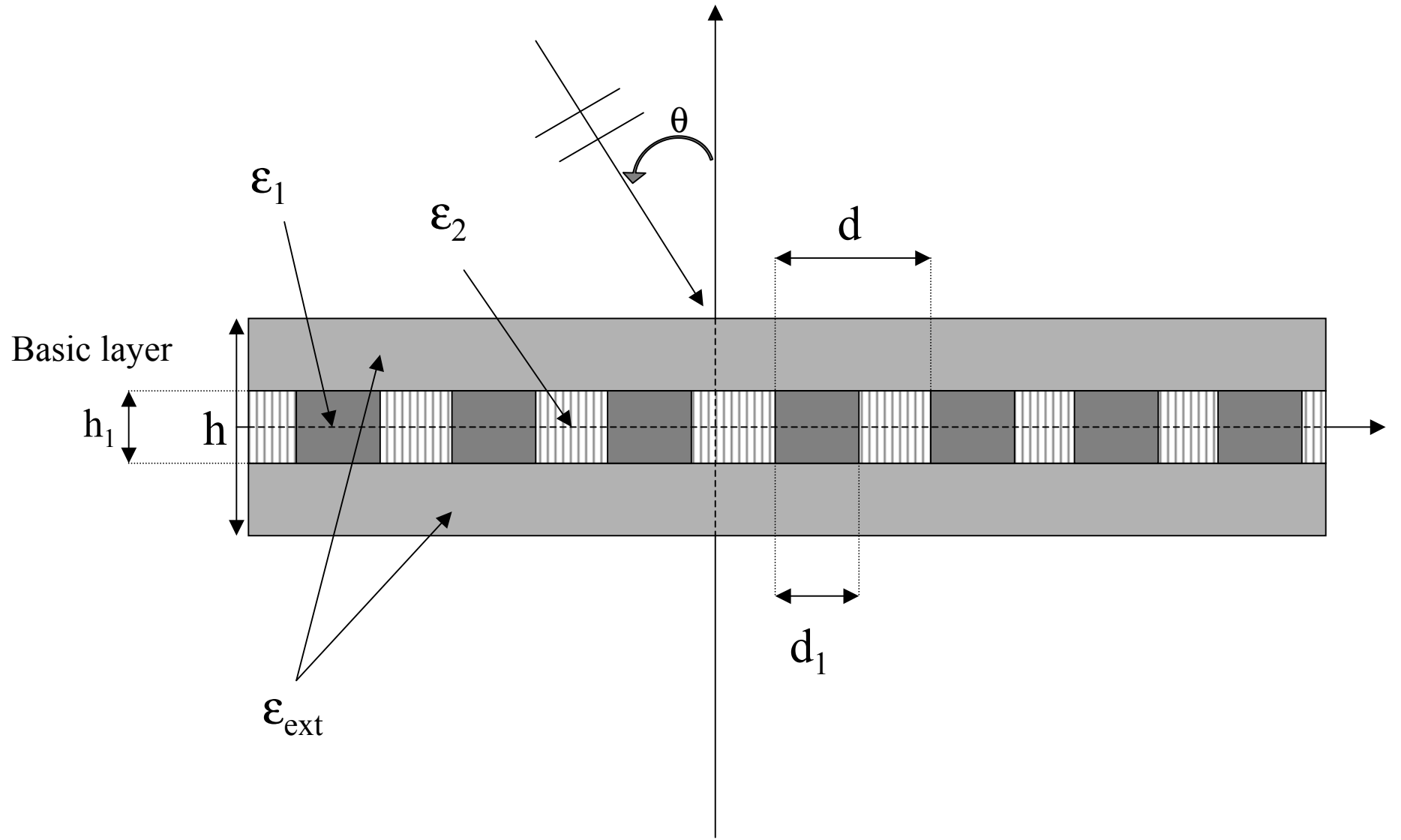
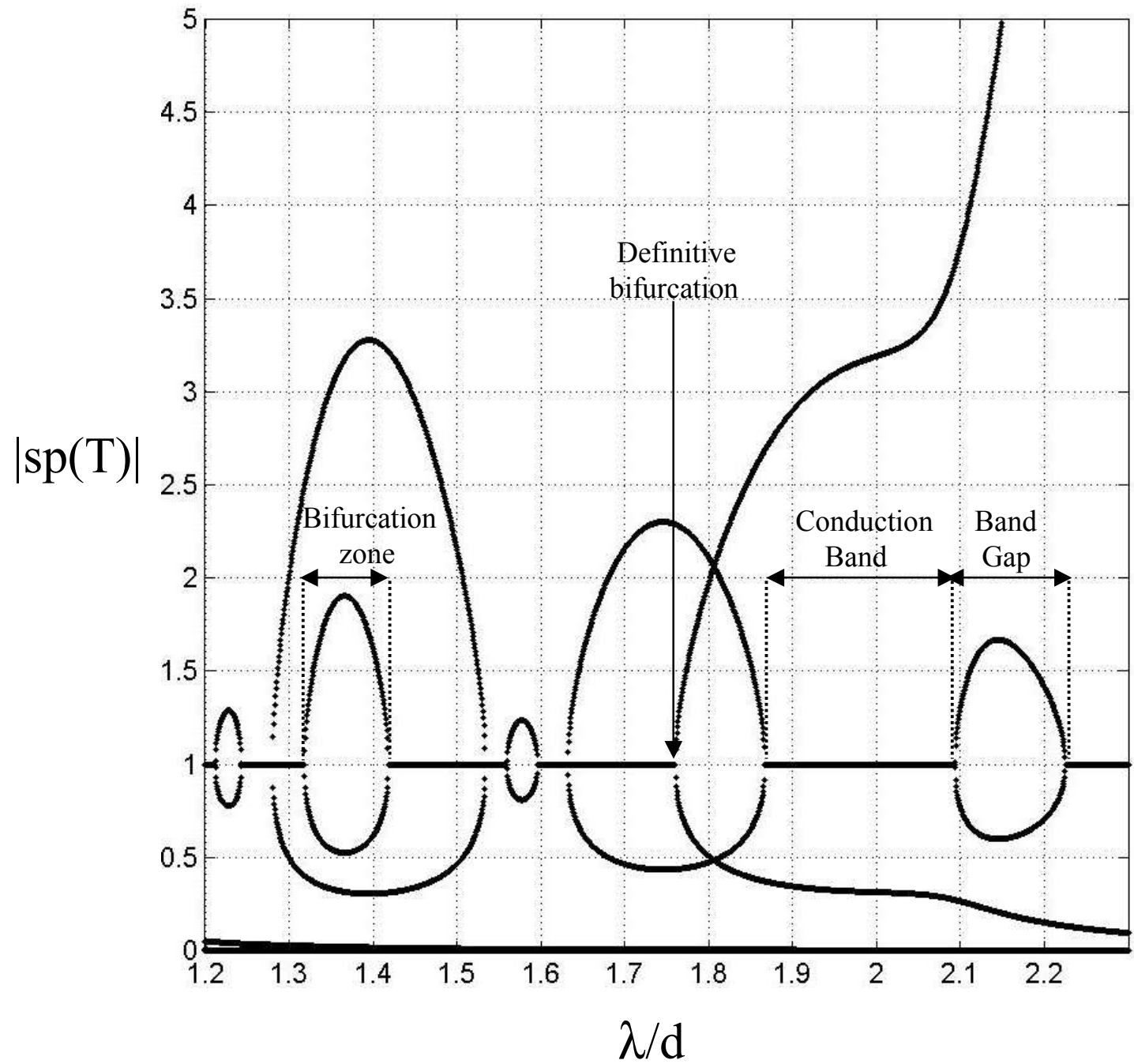
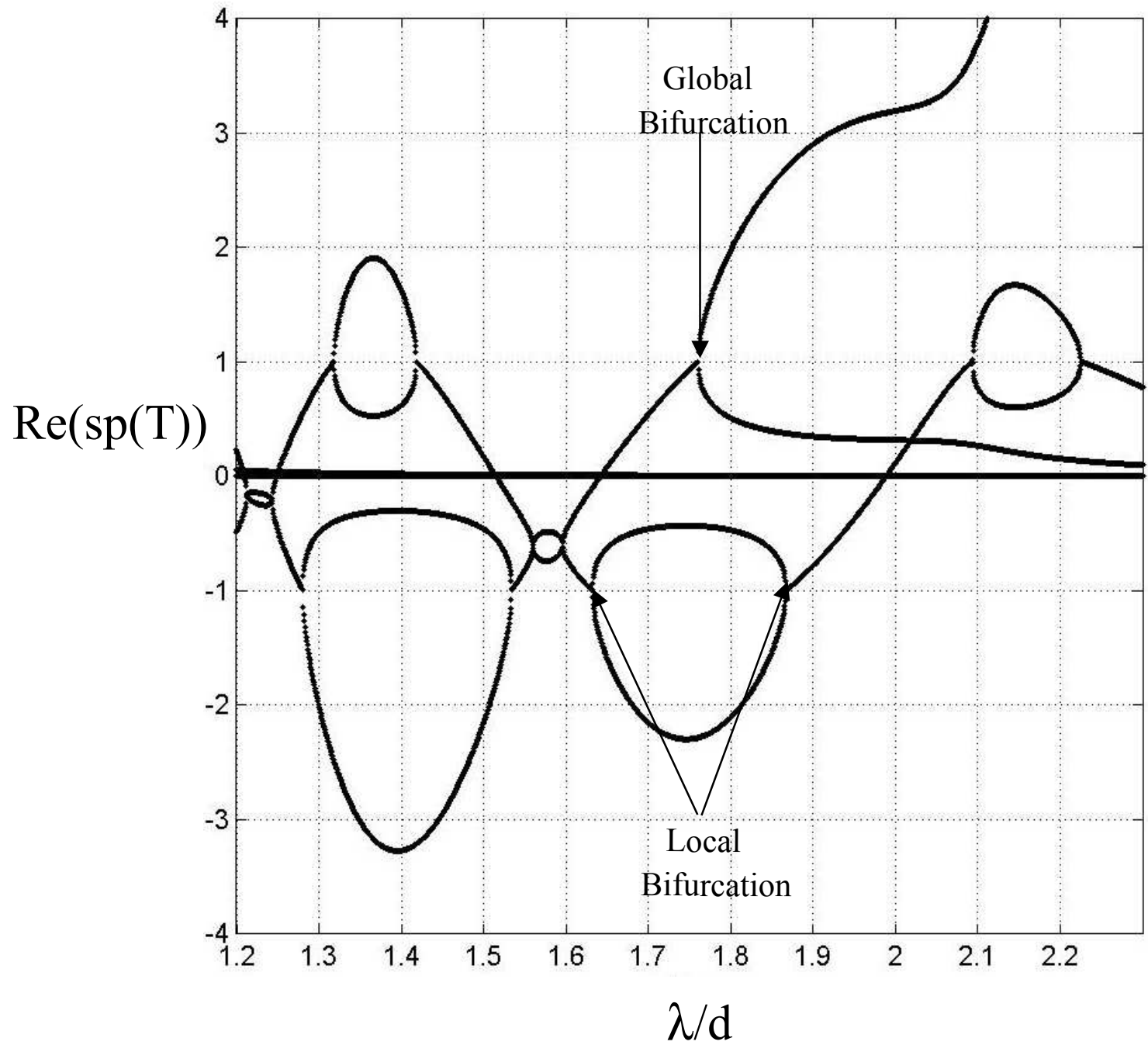
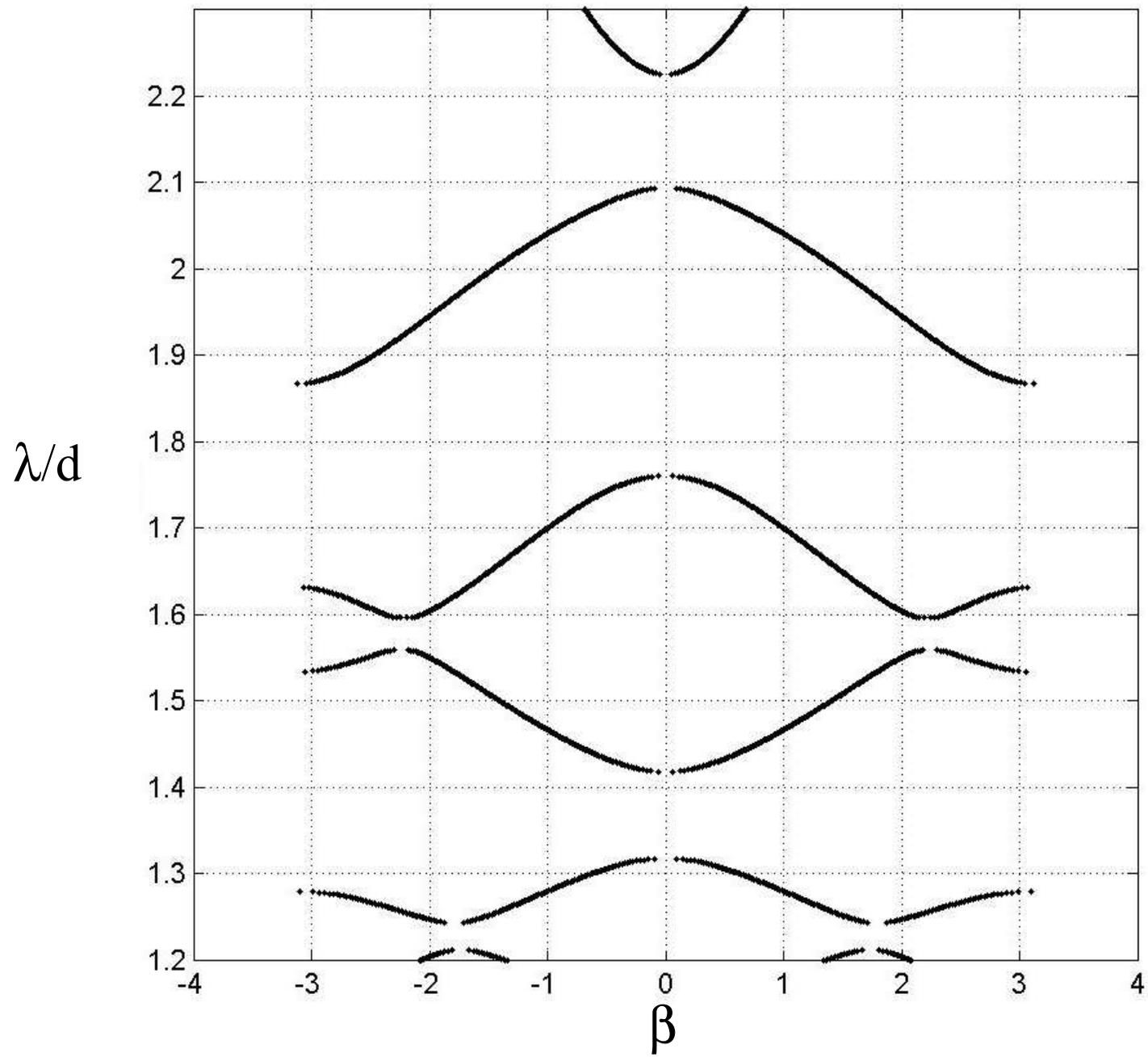
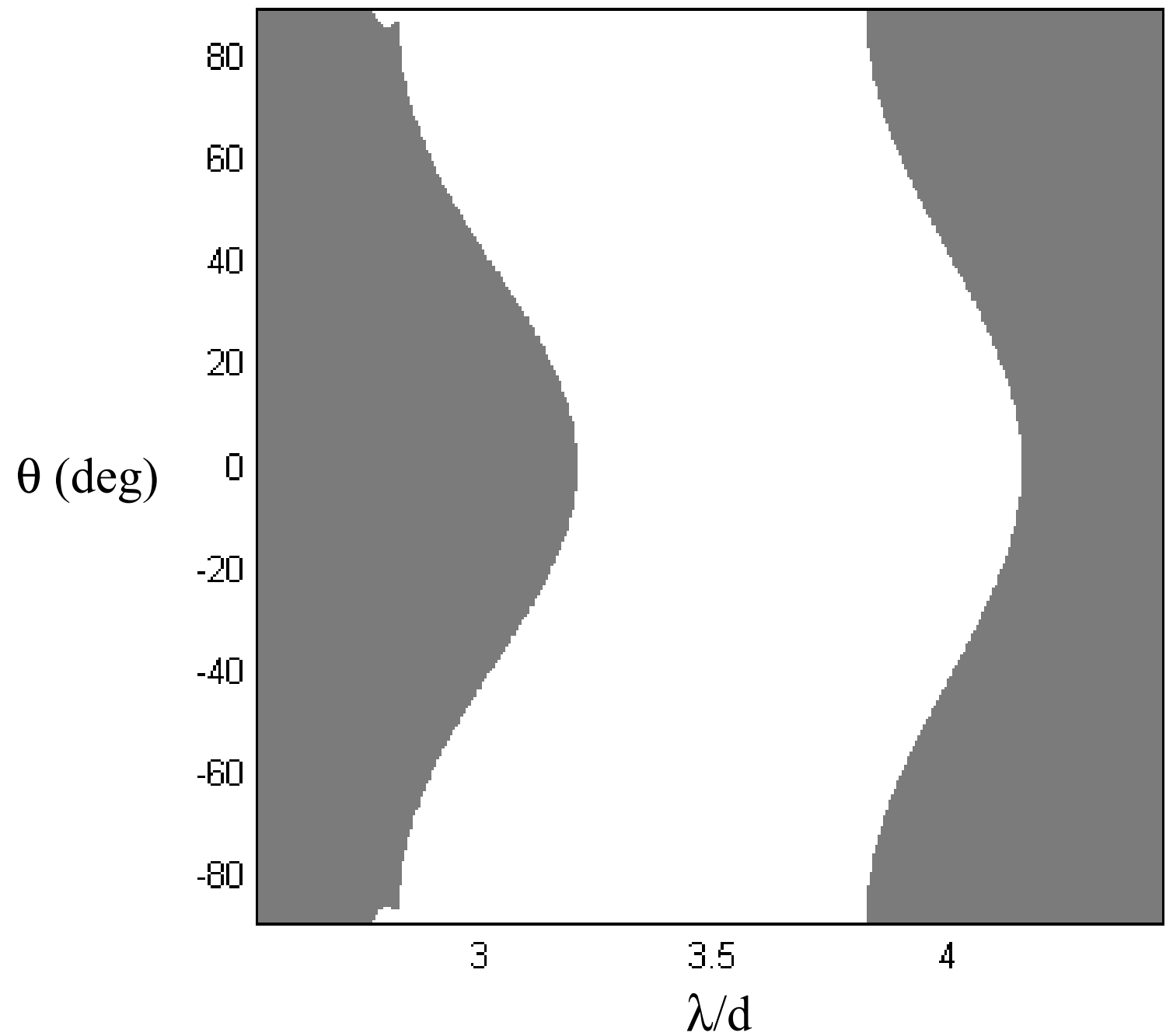
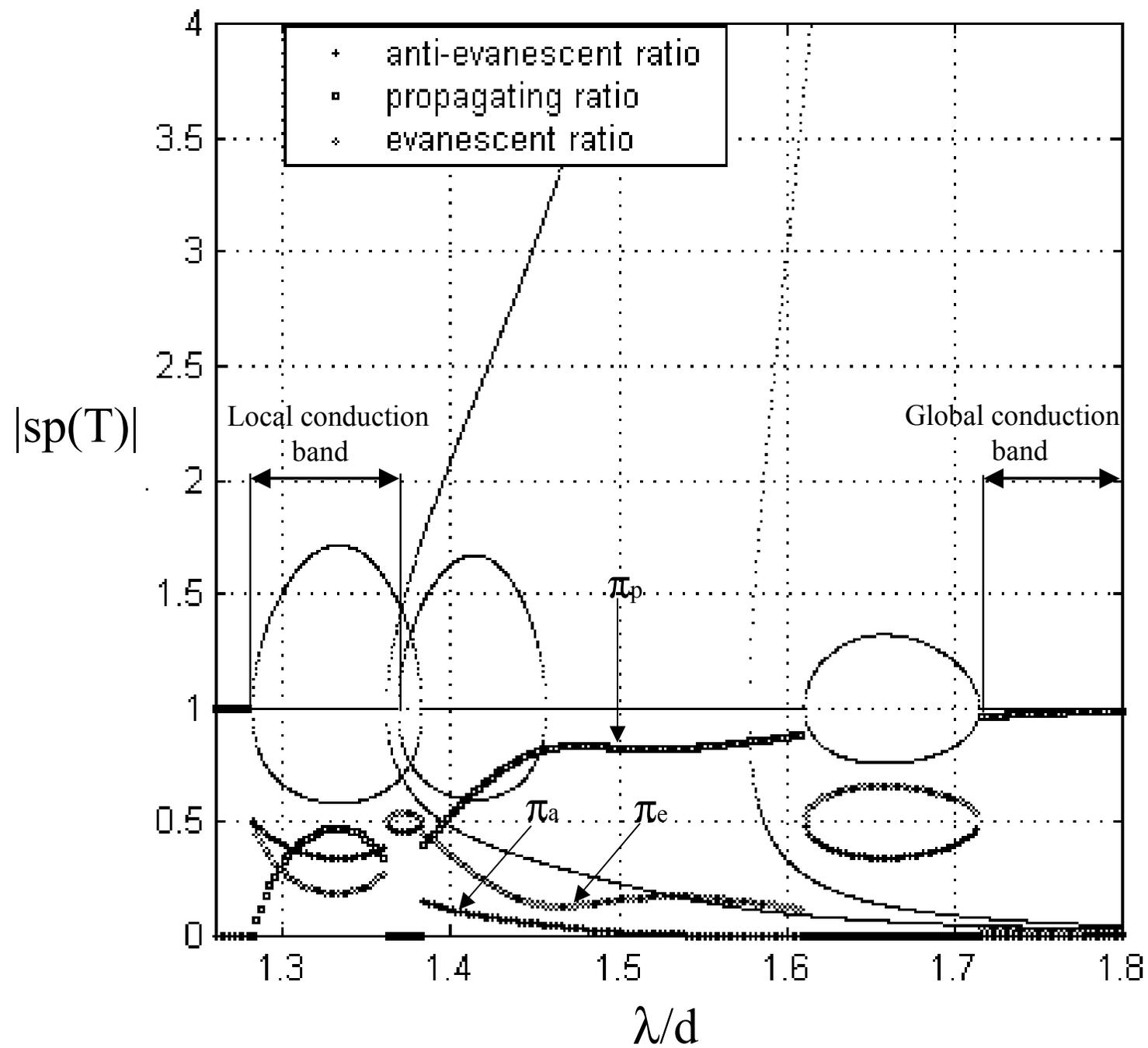
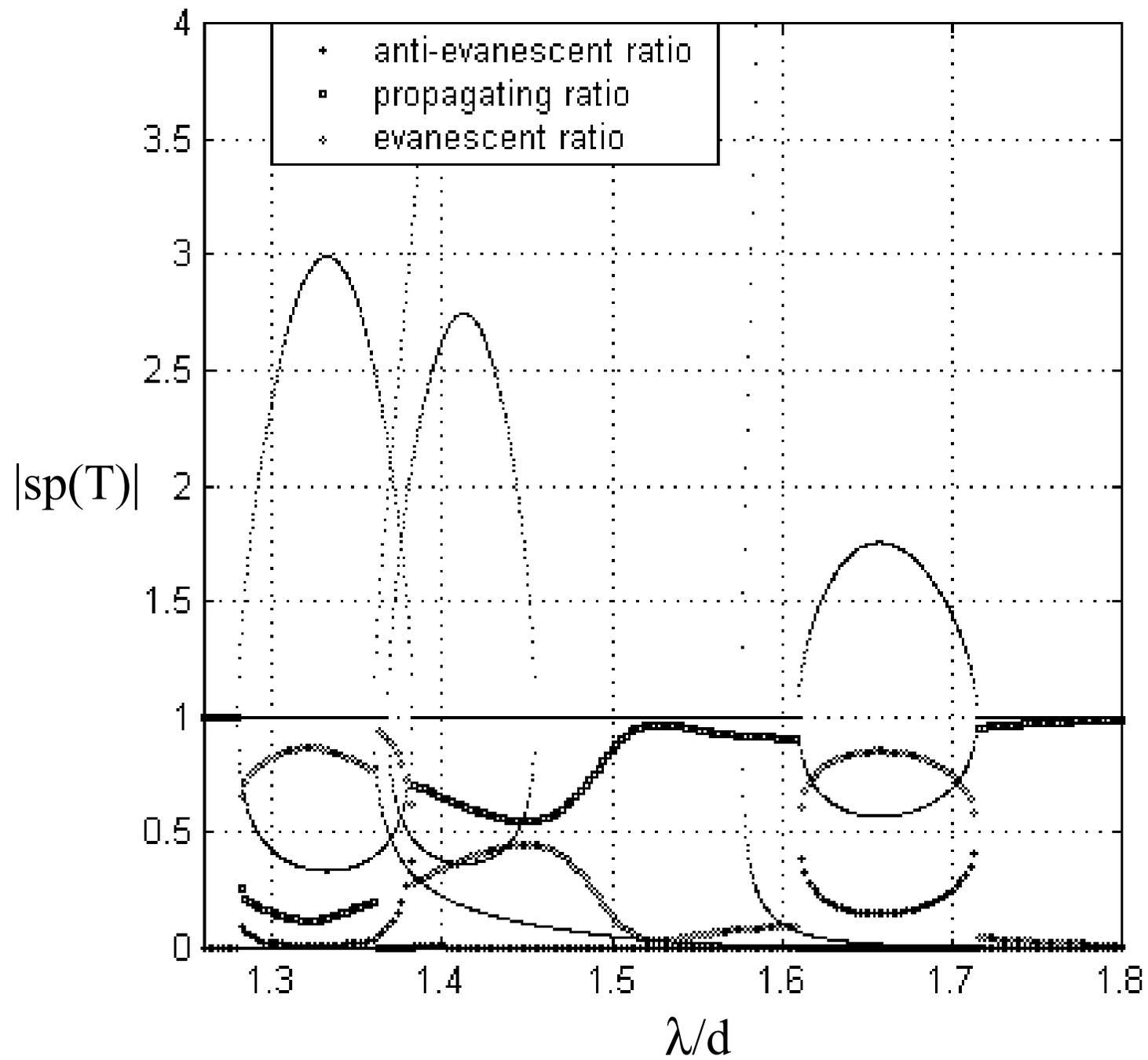
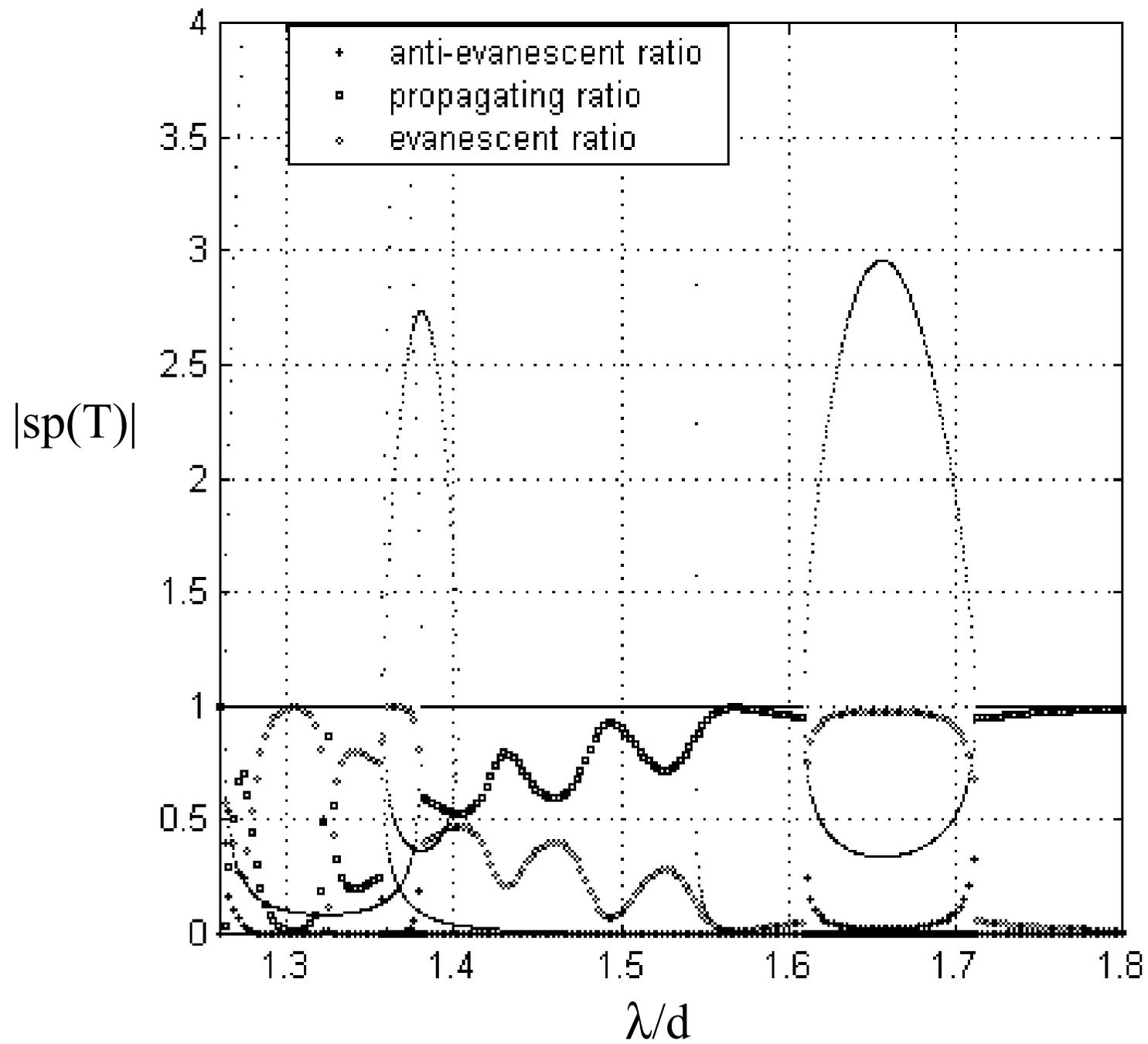
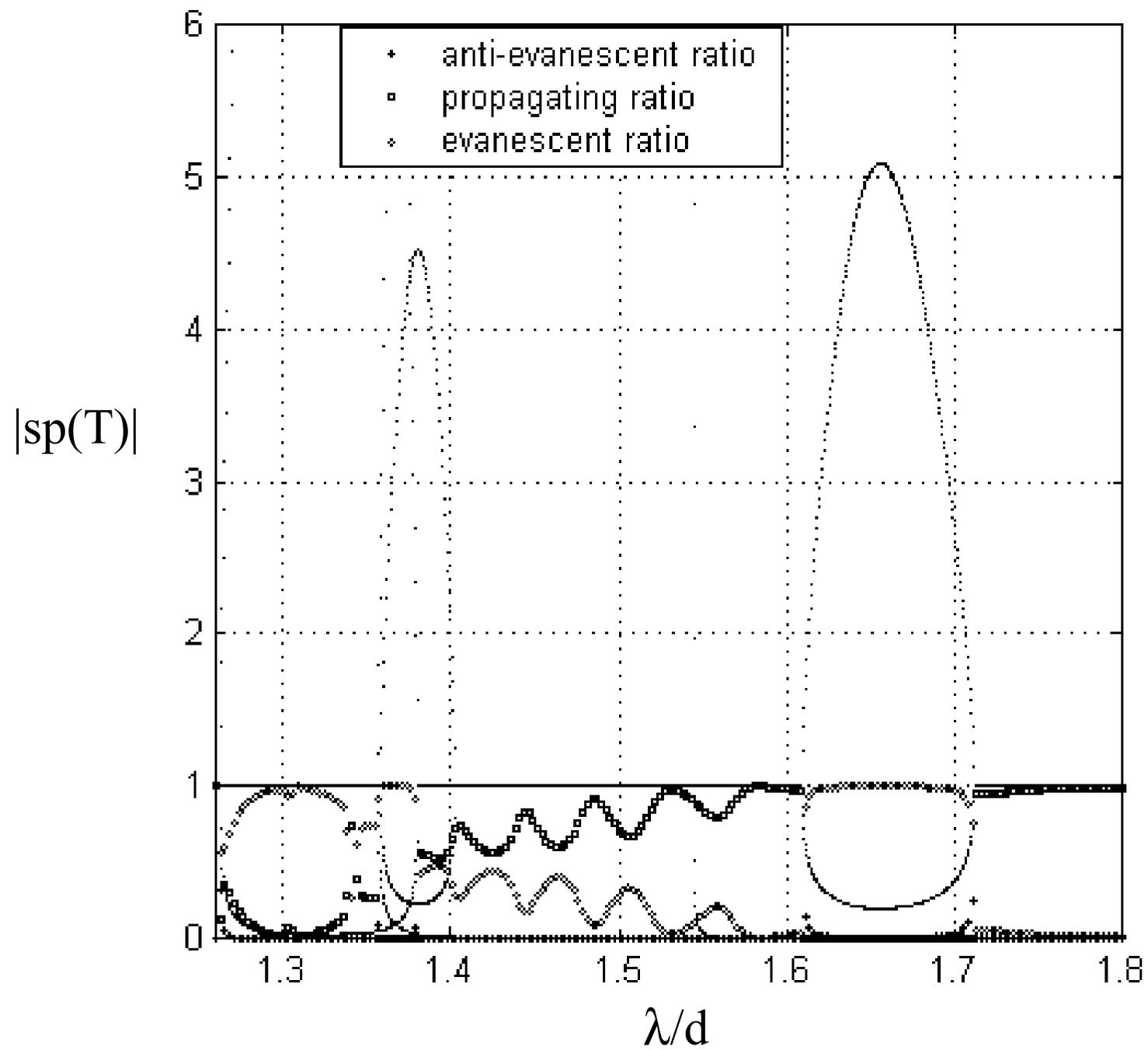
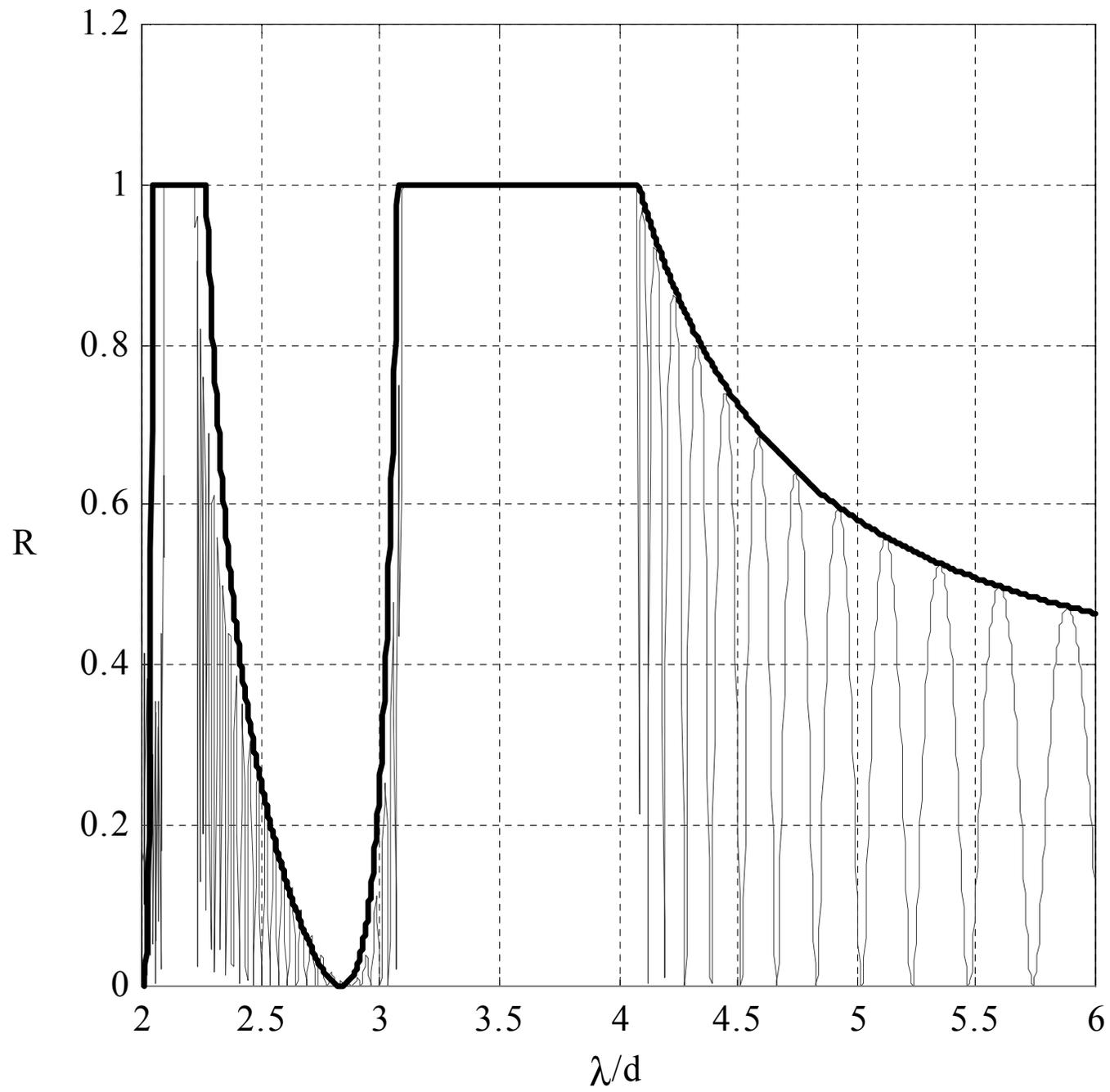
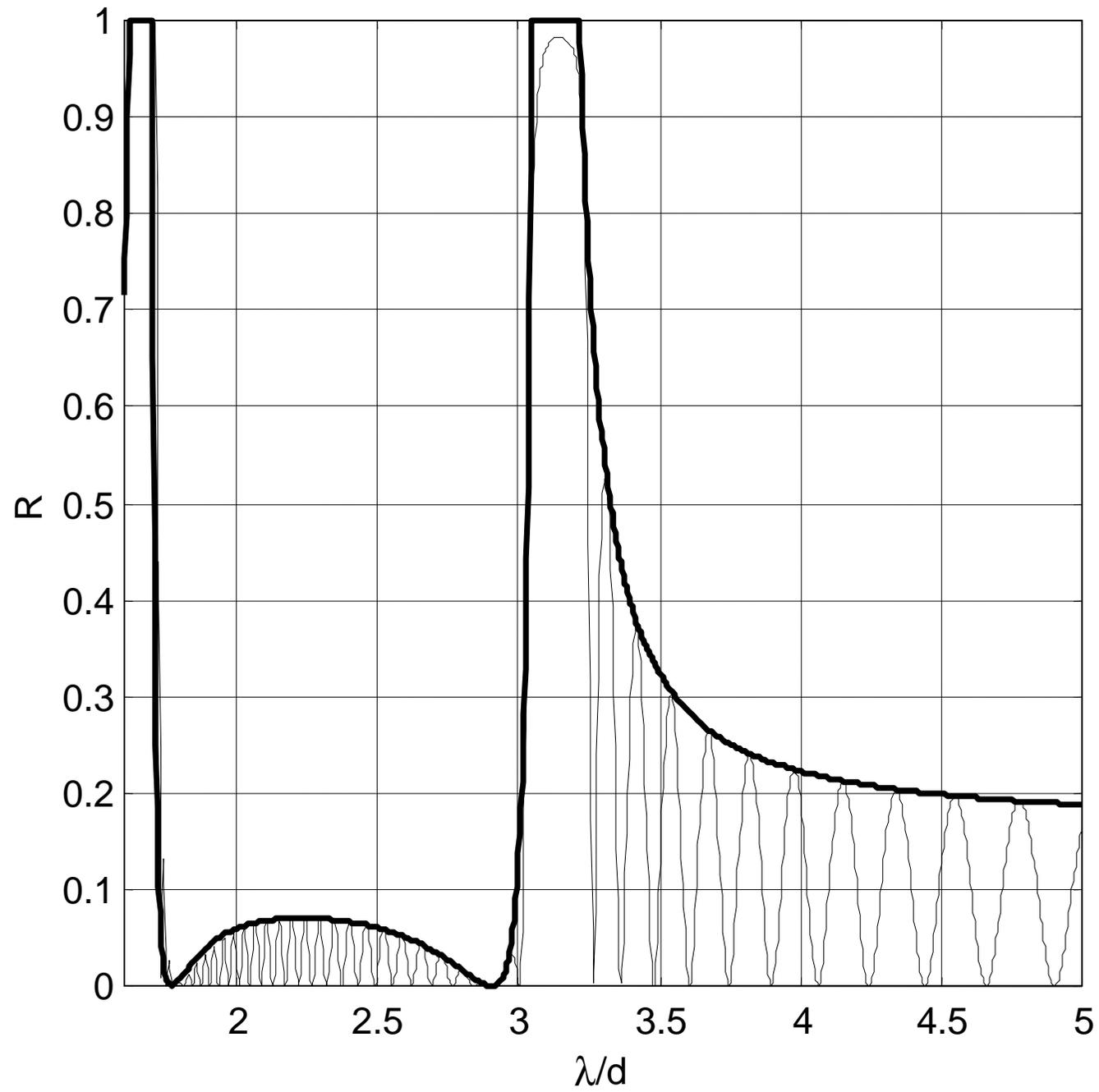
\end{document}